\documentclass[12pt,a4paper]{article}
\usepackage{graphicx}

\begin{document}
\textwidth=135mm
\textheight=200mm

\begin{center}{
{\bfseries {Search for $2K(2\nu)$-capture of $^{124}$Xe}}\footnote{
{\small Talk at The International Workshop on Prospects of Particle Physics: "Neutrino
Physics and Astrophysics" February 01 - Ferbuary 08, 2015, Valday, Russia}}

\vskip 5mm

Yu.M.~Gavrilyuk$^\dag$, A.M.~Gangapshev$^\dag$, V.V.~Kazalov$^\dag$, V.V.~Kuzminov$^{\dag,}$, \\
S.I.~Panasenko$^\ddag$, S.S.~Ratkevich$^\ddag$, D.A.~Tekueva$^\dag$, S.P.~Yakimenko$^\dag$

\vskip 5mm

{\small {\it $^\dag$ Institute for Nuclear Research, RAS, Moscow, Russia}}
\\
{\small {\it $^\ddag$ Kharkiv National University, Kharkiv, Ukraine}}
}
\end{center}

\vskip 5mm

\date{\today}%

\centerline{\bf Abstract}
The results of a search for two neutrino mode of double $K$-capture of $^{124}$Xe using a large copper low-background proportional counter are presented. Data collected during 3220 hours of measurements with 58.6 g of $^{124}$Xe provides us to a new
limit on the half-life of $^{124}$Xe regarding $2K$-capture at the level: $T_{1/2}\geq 2.0 \cdot 10^{21}$ years at a 90\% confidence level.

\vskip 5mm


\section{Introduction}

There are four types of double beta decays: double electron emission, double positron emission, single capture with single positron emission, and double electron capture (i.e. $\beta^- \beta^-$, $\beta^+ \beta^+$, $\beta^+ EC$, and $ECEC$). Double electron emission plus two $\bar{\nu}_e$ mode has been detected in direct experiments for 11 nuclei \cite{Barabash}, while the last three processes are hard for experimental study. The positive results were reported by a geochemical experiment \cite{Meshik01} for $^{130}$Ba with a half-life of $(2.2 \pm 0.5) \cdot 10^{21}$ years and a noble gas experiment \cite{PRC2013} for $^{78}$Kr with a half-life of $(9.2^{+5.5}_{-2.6}(stat) \pm 1.3(syst)) \cdot 10^{21}$ years.

For two isotopes with mass number $A = 124$ the double-beta decay is possible. The isotope of $^{124}$Sn can undergo only double electron emission. Transition of $^{124}$Xe to $^{124}$Te has a Q value of 2.866 MeV and there are three possible modes of decay: $\beta^+ \beta^+$, $\beta^+ EC$, and $ECEC$. The most likely of these three processes is $ECEC$. The main contribution to the process of $ECEC$ is ability to take two electrons from the atomic $K$-shell
\begin{eqnarray*}
^{124}{\rm{Xe}} + 2e_{K}\rightarrow ^{124}{\rm{Te}}^{**} + 2\nu_{e}.
\end{eqnarray*}

The authors of the work \cite{Mei2014} have derived upper limit of $1.66 \cdot 10^{21}$ years of the $2\nu ECEC$ process for $^{124}$Xe to the ground state of $^{124}$Te using published XENON100 experimental data \cite{XENON100}.

The data reported in Ref. \cite{Doi92} suggest that the $2K$-capture events in $^{124}$Xe take place in 76.7\%
of the total number of $ECEC$. The resulting daughter isotope $^{124}$Te$^{**}$ possess double-ionized $K$-shell.
The energy and type of emissions accompanying the process of filling these vacancies  can be evaluated, for example, on the assumption that filling a double vacancy in the $K$-shell is equivalent to simultaneous filling of two single vacancies. Each vacancy in the $K$-shell is filled by an electron from one of the above-lying shell. In this process the energy could be released with emission of Auger electrons or combination of characteristic photons (X-ray fluorescence) and Auger electrons.
The fluorescence yields for $^{124}$Te is 0.894 \cite{X-ray}.
The energies and relative intensities of the characteristic lines in the $K$ series are: $K_{ab}=31.81$~keV,
$K_{\alpha 1} = 27.47$ keV (52.2\%),
$K_{\alpha 2} = 27.2$ keV (27.7\%),
$K_{\beta 1} = 30.99$ keV (16.2\%),
and $K_{\beta 2} = 31.7$ keV (3.9\%)
\cite{X-ray}.
The total energy release being close $2K_{ab}=63.8$ keV  at simultaneous filling the double vacancy in the $K$-shell.
The probability of the emission of two characteristic X-rays and Auger electrons equal to 73.4\%.

When the process takes place in a gas phase, the characteristic photon may travel a considerable distance from the point of generation to a point of absorption. In case of events involving the nascence of two characteristic $K$-photons, the energy can be distributed in three localised regions. It is these events, which have a unique set of features, that were the subject of the search in our experiment \cite{PTE2010}. The experiment is carried out in a low-background laboratory of BNO INR RAS, at the depth of 4900 m w.e. where muon flux decreased by $10^{7}$ times in comparison with that above ground \cite{DULB}.

\section{The experimental setup}

The experimental setup consists of a large proportional counter (LPC) with a casing
made of M1-grade copper (the inner surfaces of casing and flanges were covered additionally with 1.5 and 3 mm layers of M0k copper correspondingly) and passive shield. Shield consists of 18 cm of copper, 15 cm of lead and 8 cm of borated polyethylene. The LPC is filled with the pure-xenon sample up to a total pressure of 5 bars without adding quenching or amplifying gases. In this work we use xenon enriched by $^{124}$Xe to 23\%.
Before filing, xenon is purified from electronegative  admixtures in a titanium reactor at a temperature of $600^{{\rm o}}$C.

The LPC is a cylinder with inner and outer diameters of 137 and 150 mm, respectively. A gold-plated tungsten wire of 10 $\mu$m in diameter is stretched along the LPC axis and is used as an anode. To reduce the influence of the counter edges on the operating characteristics of the counter, the end segments of the wire are passed through the copper tubes (3 mm in diameter and 38.5 mm in length) electrically connected to the anode. These segments operate as an ionisation chamber with no gas amplification. Taking into account  teflon insulators dimensions, the distance from operation region to the flange is 70 mm. The fiducial length of the LPC is 595 mm, and the corresponding volume is 8.77 $L$. The total resistance of the anode wire and two output electrodes is $\sim$ 600 Ohm \cite{PTE2010}. A detailed description of the "$2K$-CAPTURE" setup is given in Ref. \cite{PAN2013GA41}.

The detector signals are passed from one end of the anode wire to the charge-sensetive amplifier (CSA). A total shape of pulses (waveforms) are recorded by the digitizer after amplification in
an auxiliary amplifier. A digital oscilloscope LA-n20-12PCI integrated with a computer is used. Sampling frequency is 12.5 MHz and record window is 4096 bins (328 $\mu$s), $\sim 100$~$\mu$s is the "prehistory" and $\sim 228$~$\mu$s is the "history".

\section{Measurement results}

\begin{figure}[ht]
\begin{center}
\includegraphics*[width=2.75 in,angle=270.]{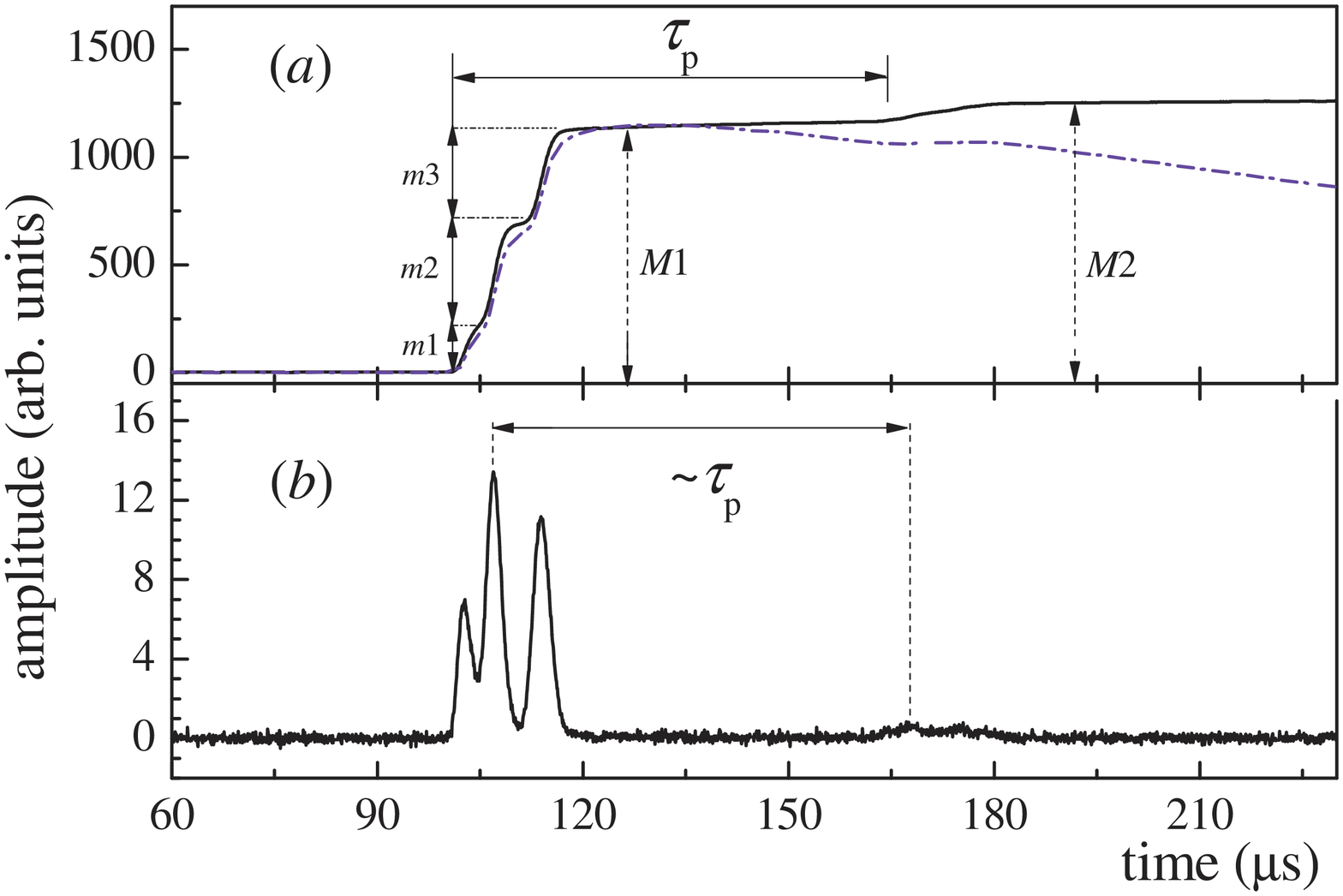}%
\caption{\label{imp}
($a$): black curve demonstrate the converted pulses of
voltage (charge) normalized for $M1$ of an original CSA pulse from three-point event (dash-dot curve).
$m1$, $m2$, $m3$ are the amplitudes of partial pulses.
The calculated, area-normalized current pulses of primary ionization electrons
are shown in $b$.
The value of $\tau_p$ is the time of the after-pulse appearance since the beginning of the primary pulse.
}
\end{center}
\end{figure}

Sample of the recorded waveform is presented in Fig.~\ref{imp}(\emph{a}).
There is a feature in the waveforms, namely the presence of secondary pulse after the primary pulse (after-pulse). Those after-pulses are result of photo emission from the cathode. Photons appear near anode wire in time of the development of an electron avalanche from the primary ionisation electrons. The relative height of the after-pulse depends on the position of the events along anode wire due to the changing of a cathode view solid angle \cite{PRC2013}. The lowest value of the angle is reached at the ends of the anode wire and the smallest secondary pulses are appeared as a result.

Offline processing of digitized pulses was performed using specially developed technique which rejects pulses of non-ionized nature \cite{PRC2013}. The signals were denoized with wavelet transformation and symmetrized by discarding ionic component contribution leaving only signal of primary electrons \cite{PTE2010}. Figure \ref{imp}(\emph{a}) show converted pulses of voltage (charge) normalized for $M1$ of an original CSA pulse (dash-dot line) obtained by integrating electric current pulses owing to primary ionization electrons (Fig.~\ref{imp}(\emph{b})) from real events which are candidates for $2K$-capture of $^{124}$Xe. The value of $\tau_p$~($\sim 63$~$\mu$s) is the time delay of the after-pulse from the beginning of the primary pulse. The amplitude of primary charge pulse $M1$ is proportional to the energy release in detector. The value of ($M2-M1$) is the amplitude of secondary charge pulse. Parameter $\lambda$-distributions
\begin{equation}\label{eq_lambda}
\lambda=(M2-M1)/M1
\end{equation}
is a relative amplitude of after-pulse which is used to select required signals. Using these boundaries we select the candidate-events for $2K$-capture from the whole set of events registered in the course of basic measurements.

\begin{figure}[pt]
\begin{center}
\includegraphics*[width=3.5 in,angle=0.]{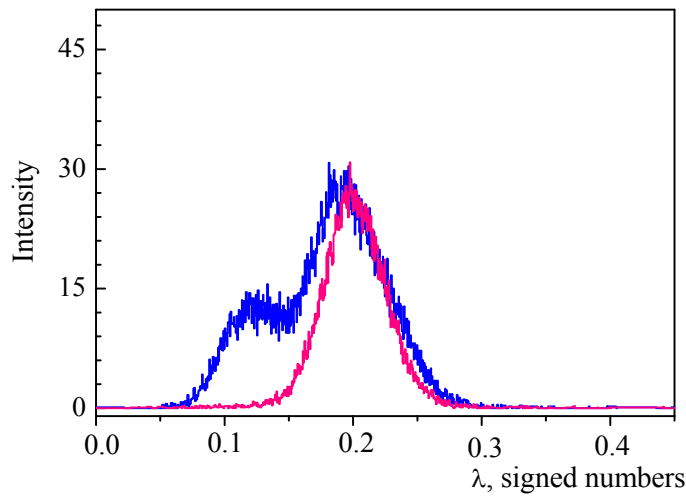}%
\caption{\label{fig_lambda}
$\lambda$-distributions of the events in the LPC filled with enriched xenon (blue curve).
The red curve is $\lambda$-distributions of the events from $^{109}$Cd source located at the middle
point of the counter length.
}
\end{center}
\end{figure}

\begin{figure}
\begin{center}
\includegraphics*[width=3.35 in,angle=0.]{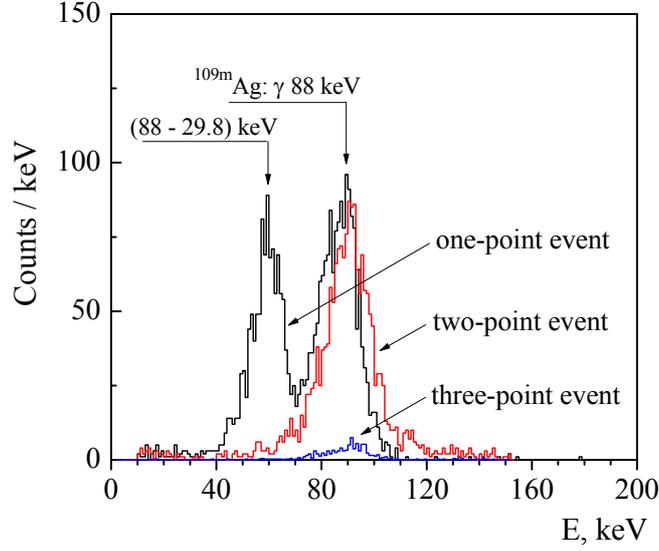}%
\caption{\label{spce_Cd}
Pulse amplitude spectrum from the irradiation with outer $^{109}$Cd  source located
at the middle point of the counter length.
$E=88-29.8=58.2$ keV is escape peak of characteristic X rays.
}
\end{center}
\end{figure}

Every $\sim2$ weeks detector was calibrated with a gamma source of $^{109}$Cd, which has an 88 keV line with 0.036 yield per decay. The photons passed through the collimating hole (12 mm in diameter and 60 mm long). $^{109}$Cd source located at the middle point of the counter length.
Figure \ref{fig_lambda}
shows the normalized distribution of the number of all events in the LPC background with enriched xenon  (blue curve) and $^{109}$Cd source (red curve), in dependence on the value of parameter $\lambda$. Comparison of the distributions shows an excess of events in the region of $\lambda < 0.18$ value, corresponding to the events from ends of the anode wire.
Figure \ref{spce_Cd} shows the pulse amplitude spectra of single-, two-, and three-point events from the irradiation with outer $^{109}$Cd source. The whole energy absorption peak shape of 88 keV line is pulled to the low energies due to the scattering in collimator and casing material.

The energy resolution (FWHM) for 58.2 keV and 88 keV lines, certain of the right half of the peak (one-point events), was found to be 16\% and 13\% correspondingly. $E=58.2=88-29.8$ keV is an escape peak, caused by
events, when on of the characteristic X-ray [$E(K_{\alpha 1})=29.8$ keV] escape the fiducial volume of the detector.

\begin{figure}[ht]
\begin{center}
\includegraphics*[width=3.25 in,angle=0.]{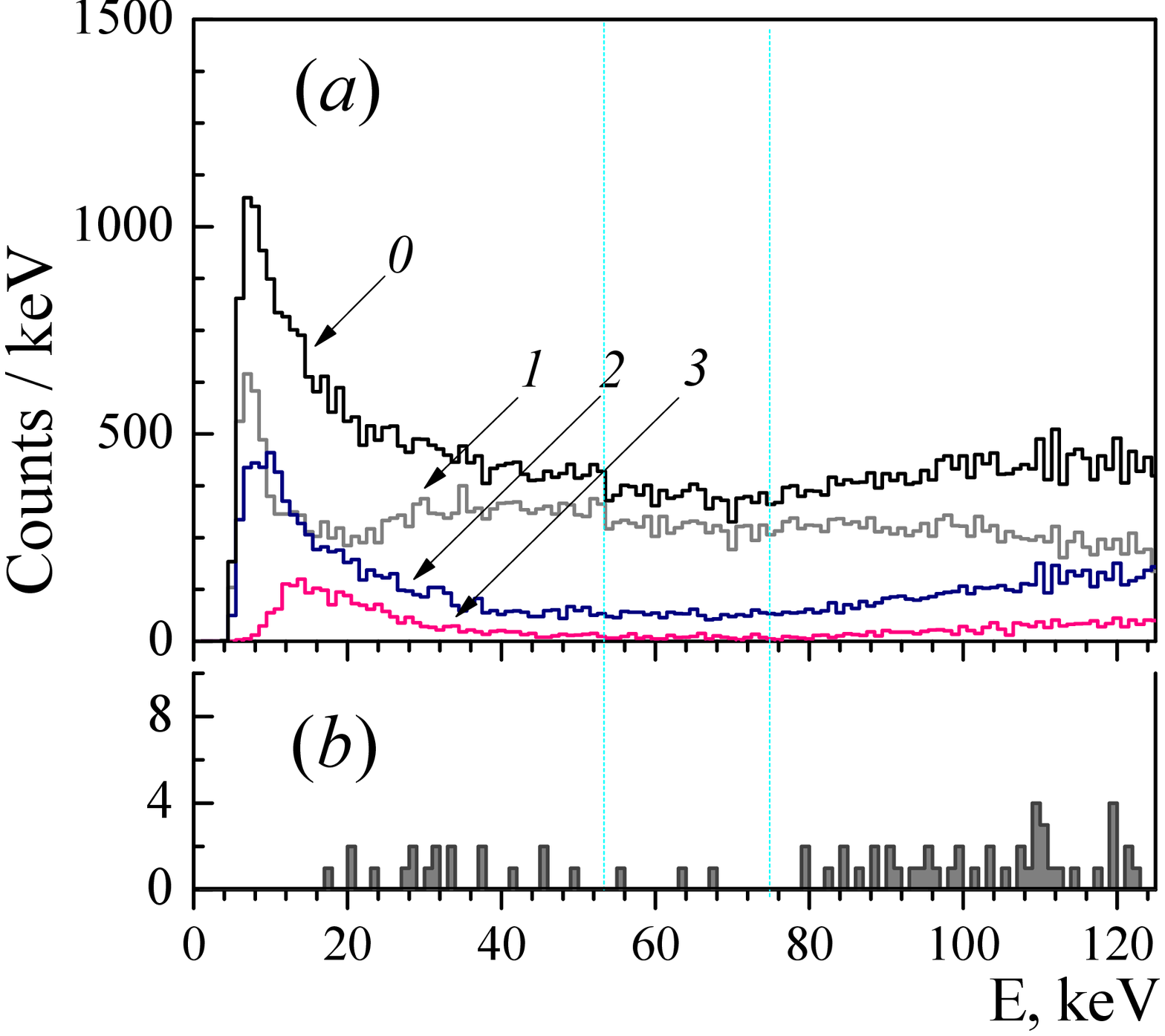}%
\caption{\label{spcA_20B40}
(\emph{a}): Amplitude spectra accumulated with the background of LPC  filled with xenon enriched in $^{124}$Xe over 3220 h
after rejection of event in parameter $\lambda$.
Curve "\emph{0}" is all events, "\emph{1}" - single-point events, "\emph{2}" - two-point events, and "\emph{3}" - three-point events.
(\emph{b}): the three-point spectra selected under the conditions "C1" and "C2".
}
\end{center}
\end{figure}

Discarding pulses for which $\lambda < 0.18$, one can fully eliminate end face events from the three-point spectra at the expense of an insignificant loss of useful events (the coefficient of useful-event selection
is $k_{\lambda} = 0.7$).
The energy spectrum accumulated with the background of LPC  filled with xenon enriched in $^{124}$Xe over 3220 h is presented in Fig.~\ref{spcA_20B40}$(a)$.
The total spectrum of all events "{\it 0}" shows dark line. Curves "\emph{1}", "\emph{2}", and "\emph{3}" give, respectively, the spectra of one-point, two-point, and three-point events.

A reduction of the background in the region of the expected effect is attained by imposing
constraints on the amplitudes of individual pulses in three-point events. For this, the amplitudes of partial pulses that arrived at the anode in the sequence $m1$, $m2$, and $m3$ are recast in the order of increase into the sequence $q0$, $q1$, and $q2$, where the amplitude $q0$ corresponds to the energy deposition from Auger electrons, while the amplitudes $q1$, and $q2$ correspond to the energy deposition of characteristic $K$ rays in the energy region of interest. With allowance for the energy resolution for the specific lines in question, the following constraints are imposed on these amplitudes: $5 \leq q0 \leq 13$ keV ("C1") and $0.7 \leq q1/q2 \leq 1.0$ ("C2"). The spectra of three-point events selected according to these criteria are given in Fig.~\ref{spcA_20B40}$(b)$.

The background in  the required energy region from 52 to 78 keV in this spectrum was three events over $t_{meas}$=3220 h of measurement. Assuming absence of background events in ROI we get a conservative upper limit on counts from double $K$-capture of $^{124}$Xe at the level n$_{exp}$=7.42 according to the Feldman-Cousins procedure \cite{FC}.

Half-life limit is set to the two neutrino double-$K$-electron capture process for $^{124}$Xe to the ground state of $^{124}$Te. The half-life limit has been calculated using formula
\[
\mathrm{lim}
\begin{array}{*{3}c}
   {} \\
\end{array}
T_{1/2}  = ln2 \cdot N_{124} \cdot t_{meas} \cdot \frac{p_3 \cdot \varepsilon_p \cdot \varepsilon_3 \cdot \alpha_k \cdot k_\lambda}
{n_{exp}},
\]
where $N_{124}=2.85\times10^{23}$ is the number of $^{124}$Xe atoms in the operating volume of the counter, $p_3=0.809$ is the fraction of $2K$-captures accompanied by the emission of two \emph{K} X-rays; $\varepsilon_p=0.735$ is the probability of two \emph{K}-photon to be absorbed in the operating volume; $\varepsilon_3=0.51$ is the efficiency to select three-point events due to $2K$-capture in $^{124}$Xe; $\alpha_k=0.985\pm0.005$ is the fraction of events with two \emph{K}-photon that could be registered as distinct three-point events; $k_\lambda=0.7\pm0.1$ is the useful event selection coefficient for a given threshold for $\lambda \geq 0.18$.

The obtained result is
\[
    T_{1/2}^{2\nu 2K} (g.s.\rightarrow g.s.) \geq 2.0\times 10^{21}\texttt{yr}~(90\%~\texttt{C.L.}).
\]

\section{Conclusions}
An experiment to search for double beta decay processes in $^{124}$Xe (58.6 g)
with the help low background proportional counter
is in progress at the "$2K$-CAPTURE" setup of the underground laboratory DULB-4900 of BNO INR RAS \cite{DULB}.
The sensitivity of the experiment after 3220 h of data taking is on the level of lim $T_{1/2} \sim  10^{21}$ yr for the double $K$-capture processes in $^{124}$Xe with emission of two characteristic photons ($K_{\alpha_1}$ and $K_{\alpha_2}$) and Auger electrons.

We hope to improve further the sensitivity of the experiment for this channel of the decay to the level of theoretical predictions
($T_{1/2} \sim 7.3\cdot10^{21} - 10^{22}$ yr \cite{Hirsch94, Urin98, Singh2007, Suhonen2013}) by increase of the statistics  and more thorough pulse-shape discrimination.

This work is supported by the Program of basic researches of the Presidium of the Russian Academy of Sciences "Fundamental properties of matter and astrophysics" for the support of the research.

\end{document}